# Compressive Behavior and Failure Mechanisms of Freestanding and Composite 3D Graphitic Foams

Kenichi Nakanishi,[†] Adrianus I. Aria,[‡] Matthew Berwind,[∥] Robert S. Weatherup,[∇, §] Christoph Eberl,[∥] Stephan Hofmann,[†,*] and Norman A. Fleck[†,*]

[†] Department of Engineering, University of Cambridge, Cambridge, United Kingdom CB2 1PZ

[‡] Surface Engineering and Nanotechnology Institute, School of Aerospace, Transport and Manufacturing, Cranfield University, Cranfield, United Kingdom, MK43 0AL

[∥] Fraunhofer Institute for Mechanics of Materials IWM, Freiburg, Germany 79108

[∇] School of Chemistry, University of Manchester, Oxford Road, Manchester M13 9PL, UK

[§] University of Manchester at Harwell, Diamond Light Source, Didcot, Oxfordshire, OX11 0DE, UK

**Abstract**

Open-cell graphitic foams were fabricated by chemical vapor deposition using nickel templates and their compressive responses were measured over a range of relative densities. The mechanical response required an interpretation in terms of a hierarchical micromechanical model, spanning 3 distinct length scales. The power law scaling of elastic modulus and yield strength versus relative density suggests that the cell walls of the graphitic foam deform by bending. The length scale of the unit cell of the foam is set by the length of the struts comprising the cell wall, and is termed level I. The cell walls comprise hollow triangular tubes, and bending of these strut-like tubes involves axial stretching of the tube walls. This length scale is termed level II. In turn, the tube walls form a wavy stack of graphitic layers, and this waviness induces interlayer shear of the graphitic layers when the tube walls are subjected to axial stretch. The thickness of the tube wall defines the third length scale, termed level III. We show that the addition of a thin, flexible ceramic $Al_2O_3$ scaffold stiffens and strengthens the foam, yet preserves the power law scaling. The hierarchical model gives fresh insight into the mechanical properties of foams with cell walls made from emergent 2D layered solids.

Keywords: Cellular solids, Chemical vapor deposition (CVD), Graphene, Micromechanical modeling, Structural hierarchy

* Corresponding authors: naf1@eng.cam.ac.uk, sh315@cam.ac.uk

1. Introduction

Recent advances in manufacturing methods for 2D materials, and in particular graphene, allow these layered solids to be engineered as the cell walls in a new class of porous cellular materials [1–10]. The ability to tailor unique combinations of mechanical, thermal, electrical and optical properties and to achieve ultra-low density and high surface area gives the opportunity for these 2D solid-based cellular materials to find applications ranging from (opto)electronics, artificial skin, robotics, electrochemistry, and catalysis to thermal management, self-cleaning, sorption and filtration, and biomedical devices [11–16]. The current literature on the mechanical response of foams is focused on cell walls made from conventional metals, polymers, and ceramics [17–27]. An understanding of the mechanical properties of 2D materials-based cell walls is required for all of these emerging applications.

The microstructure of the constituent graphene/graphite in porous materials can vary significantly. There are two general classes of microstructure for graphene-based foams, platelets versus sheets. Aerogels comprise platelets at the cell wall level, with a flake diameter typically on the order of tens of μm [28,29]. These flakes depend upon weak inter-flake interactions [30–32], and unless reinforced by a binding agent, mechanical tests can result in flake debonding and fragmentation. In contrast, continuous multi-layer graphene films with high integrity can be made via chemical vapor deposition (CVD), which relies on a 3D template that is exposed to reactive gases at elevated temperatures to crystallize 2D materials on its surface [12,33–35]. This enables the creation of macroscopic foams endowed with a covalently bonded, poly-crystalline network structure, and the possibility of improved macroscopic properties.

Graphitic multi-layers have a very low out-of-plane shear stiffness and strength compared to their exceptional in-plane modulus. Consequently, it is unclear whether foams made from CVD graphene will possess high stiffness and strength (as dictated by the in-plane properties) or a much lower stiffness and strength due to shear deformation. This competition has not been addressed before within either the foam literature or graphene literature. One of the few recent mechanical studies [36] on freestanding CVD graphitic foam reports a macroscopic Young's modulus of 340 kPa at a relative density of 0.002. This low value was attributed to a high defect density and bending of the cell walls of the foam. Here, we explore the compressive response of freestanding CVD graphitic foams, using a multi-scale modeling framework. We also explore the effect of a thin $Al_2O_3$ scaffold, deposited by atomic layer deposition (ALD), upon the macroscopic properties of such foams. The present study is an investigation of a new class of foams, with the potential for a paradigm shift in performance.

## 2. Manufacture: Methods and Resulting Microstructure

*2.1 Manufacture of free-standing graphitic foams*

Processing methods have recently evolved for CVD-based graphitic foams, and much of the recent literature is based on commercial Ni foams [11,12,34,37–40]. The quality and thickness of the CVD graphitic multi-layers can be controlled by growth parameters such as time, temperature, precursor concentration and catalyst choice [41]. For such freestanding graphitic foams to be mechanically stable, the walls are grown to multi-layer (>30) thickness. The microstructure of the CVD graphitic walls is thereby polycrystalline, with properties that depend upon the level of graphitization, homogeneity and defect density.

Consider route (i) of Figure 1. 3D graphitic foams (GF) were prepared by CVD on sacrificial open-cell Ni foam templates, with a purity of >99.99%, 95% porosity, 1.6 mm thickness and a bulk density of 450kg/m$^3$. The CVD process was carried out at a total pressure of 50 mBar, combining diluted $CH_4$ (5% in Ar) and $H_2$ in a 100 mm hot-walled tube furnace at a temperature of 950°C for 3 hours, as described in detail elsewhere [42]. A $CH_4$:$H_2$ ratio of 1:1 to 3:1 was used during the growth; a richer hydrocarbon atmosphere resulted in a higher average number of graphene layers and thus a higher relative density. Samples were cooled at a maximum rate of 20 °C/min after growth. CVD grown samples had dimensions of 25 mm x 80 mm x 1.6 mm and were stored in ambient conditions prior to further processing and characterization.

Polycrystalline graphitic layers grow in parallel to the Ni surface for all grain orientations, resulting in continuous sheets that envelop the catalytic template [43]. This templating effect of a Ni catalyst in the growth of 2D carbon materials is well known: where graphitic lattice-metal interactions at the nanoscale dictate the nucleation and growth dynamics and thus the final lattice morphology [41,44]. High-resolution transmission electron microscopy (HR-TEM) of a cross-section of a CVD graphitic film deposited on the surface of a Ni layer shows graphitic layers running parallel to the metal surface with characteristic (002) graphite spacing, see Fig. 2. Other HR-TEM studies have confirmed the formation of covalent bonds at grain boundaries, allowing the in-plane Young's modulus to remain high [45,46].

After the graphitic layer CVD, the sacrificial Ni templates were removed using wet chemical etching. To prevent collapse of the graphitic foam structure by capillary condensation, a PMMA scaffold was used. Samples were dip coated in PMMA (495K, 2% in anisole) for 10 s, then annealed on a hot plate at 180°C for 15 minutes on each side, trimmed along the edges and then etched in 0.5M $FeCl_3$ for 48 hours. Subsequently, the samples were rinsed in de-ionized water, then subjected to a 30 minute etch in 10% HCl solution to remove Fe residue. Finally, samples were given a wash in de-ionized water, and dried in ambient air. At this stage, the graphene foam samples are metal-free but remain supported by the PMMA scaffold. The PMMA supporting structure was removed by annealing the sample at 450°C for 60 minutes in a $H_2$/Ar (1:5 ratio, 50 mBar) atmosphere to produce free-standing graphitic foams (FG) [47,48].

*2.2 Manufacture of alumina supported graphitic foams*

Consider route (ii) of Figure 1. As before, the CVD of graphitic layers was carried out using the Ni foam template. Aluminum oxide ($Al_2O_3$) coatings were subsequently deposited onto the graphitic layers by ALD in a multi-pulse mode at 200°C, as described in detail elsewhere [49]. Trimethylaluminum (TMA, purity >98%) was used as the precursor and deionized water vapor ($H_2O$) as the oxidant.  $Al_2O_3$ layers of 50 nm thickness were deposited by applying 550 deposition cycles. Then, the original Ni template was removed as in Section 2.1 to produce free-standing, alumina supported, graphitic foams ($Al_2O_3$/G).

*2.3 Structural and Elemental Characterization*

All samples were characterized by scanning electron microscopy (SEM) at a 2kV accelerating voltage. Raman spectra were measured at room temperature using a 532 nm wavelength laser with a 50x objective. Exposure times of less than 2s were used to avoid detrimental laser heating of the specimens. Elemental analysis of the samples was performed using X-ray photoelectron spectroscopy (XPS) at an operating pressure below $10^{-10}$ mbar. The X-ray source for the XPS was a monochromated Al Kα with a photon energy of 1486.6 eV and a spot size of 200 µm. All XPS spectra were acquired from the internal walls of laser-cut sample cross-sections to ensure that the collected spectra accurately represent the hollowed sample. Bright-field HRTEM images were collected at an accelerating voltage of 400 kV. Thermogravimetric analysis (TGA) was carried out in synthetic air (20% $O_2$ in $N_2$). A 2 µg portion of each sample was ramped from room temperature to 900°C. During the measurement, the temperature was held for 15 min at 100°C to completely remove adsorbed water.

Compression tests were performed on samples of height 1.6 mm and cross-section 5 mm x 5 mm. All samples were laser cut to the desired dimensions to ensure that faces were parallel. The density of each sample was deduced by measurement using a high-precision electronic balance. A custom-built mechanical testing apparatus was used to measure the compressive response of the laser-cut foam samples. This system consists of a stepper-motor driven linear actuator for positioning (50 nm resolution) and a preloaded piezoactuator (1.2 nm resolution) for displacement actuation. A miniature tension/compression load cell was used for force measurement (5 N range and a 2.5 mN resolution). Each specimen was aligned along the loading axis of the test system and fastened electrostatically to a parallel aluminum plate base. All experiments were conducted at a nominal displacement rate of 10 µm/s, implying a strain rate of $1.6 \times 10^{-4}$ $s^{-1}$ [50], and the deformation of the microstructure was observed by in-situ microscopic image acquisition.

*2.4 Microstructure*

SEM images of the as-fabricated freestanding graphitic (FG), and alumina supported ($Al_2O_3$/G) foams, are shown in Figure 3. The unit cell length (denoted *L* in Fig. 3a) is in the range 200–400 µm for both FG and $Al_2O_3$/G (Fig. 3a). The hollow struts have triangular cross-sections with side lengths of d = 30–70 µm for both FG and $Al_2O_3$/G (Fig. 3b). This large variation in the value of *d* is inherent to the commercial open-cell Ni templates that are used herein [18,51]. The cellular geometries of the foams are neither altered by the CVD process nor by the Ni removal, as seen by comparing Figure 3 to the SEM of the original Ni foam template (see Supplementary Material Fig. 3). The thickness of the strut walls (denoted by *h*) is measured from SEM images of the cross-section (Fig. 3c). For FG, *h* equals 80-150 nm and the relative density $\bar{\rho}$ corresponds to 0.002-0.005 (Fig. 3c). This range of relative density and associated wall thickness is comparable to values reported in the literature for device applications [12,33,36,52,53]. For the ceramic ALD coating, we focus on a fixed $Al_2O_3$ thickness of 50 (±5) nm (Fig. 3c), which is sufficiently thin for the alumina to remain flexible [54], but sufficiently thick to give a measureable change in the macroscopic compressive properties. Hence, for $Al_2O_3$/G samples, h ranges from 130 nm to 200 nm (Fig 3c).

It was not practical to construct FG structures of *h* below 30 nm ($\bar{\rho} \leq 0.001$) as they were not stable upon removal of the Ni template; they are easily damaged by electrostatic forces, making sample handling and reproducibility in mechanical measurement impractical.

Graphitic foams of relative density below 0.001 are generally supported by a polymeric layer [55–57] due to such problems of instability and irreproducibility.

The strut walls exhibit waviness throughout the volume of the foam (Fig. 3b). The length scales of wall waviness are obtained via SEM imaging of the surface of the strut walls (see Supplementary Material Fig. 6). The line profiles of the wall surfaces indicate a characteristic variation on the scale of a few μm. We idealize this roughness as a sine wave, with a characteristic amplitude $w_0$ = 0.76 - 2.8 μm and wavelength $\lambda$ = 3.7 – 18 μm for both FG and $Al_2O_3$/G (Table 1). We propose that the waviness relates to the polycrystalline grain structure of the commercial Ni foams (see Supplementary Material Fig. 3b), for which grains range in initial size from 4-20 μm, and the presence of multiple different Ni surface orientations as a result of the non-planar shape of the foams, leading to inhomogeneities during CVD of the graphitic layers, see Fig. 1a.

### 3. Mechanical Characterization

*3.1. Compression tests*

A typical plot of nominal compressive stress σ versus nominal (engineering) strain ε for FG is given in Fig. 4, for a displacement rate of 10 μms$^{-1}$. As noted for a wide range of foams [25], three distinct regimes exist: (I) linear elastic for ε less than the yield strain $\varepsilon_Y$, (II) plateau $\varepsilon_Y$ < ε < $\varepsilon_D$, where $\varepsilon_D$ is a densification strain and (III) densification ε > $\varepsilon_D$, (Fig. 4a). In regime I, the foam is strained in a uniform elastic (i.e. reversible) manner, with no observable damage evolution. The onset of plasticity marks the change from regime I to regime II. There is a clear change in slope in Figure 4a at the onset of plastic collapse (at ε = $\varepsilon_Y$).

In order to obtain insight into the collapse mechanism, a specimen was subjected to successively larger levels of macroscopic strain ε, followed by unloading to zero load and the remnant strain $\varepsilon_r$ was measured from the associated SEM images, see Fig. 4b. A series of dotted lines are shown in Fig. 4a to give the end points of this elastic unloading. These images reveal the following:

(i) Straining is elastic up to the onset of plastic collapse ($\varepsilon = \varepsilon_Y = 0.14$) such that $\varepsilon_r = 0$. For example, full recovery is observed from an imposed strain level of $\varepsilon = 0.1$, (point B1) as shown in Fig. 4a.

(ii) After straining to a level $\varepsilon > \varepsilon_y = 0.14$, the foam exhibits plastic collapse with little observable microcracking or debonding of the struts. For example, elastic unloading from $\varepsilon = 0.4$ (point C1) results in a remnant strain $\varepsilon_r = 0.24$ (point C2).

(iii) When the specimen is strained to beyond a densification strain $\varepsilon_D = 0.38$, the struts impinge upon each other and strong strain hardening occurs. The full unloading curve from $\varepsilon = 0.6$ (point E1), to $\varepsilon_r = 0.46$ (point E2) is also shown in Fig. 4a and reveals a non-linear unloading behavior associated with the elastic relaxation of the distorted microstructure as the strain reduces to $\varepsilon_r$.

A higher resolution image of the deformed struts after unloading from $\varepsilon = 0.6$ is shown in Fig. 5a. Plastic hinges are marked by the arrows in Fig. 5a, the formation of which is schematically shown in Fig 5b. The resemblance between the deformed microstructure of the graphitic foam and of Ni INCOFOAM® in compression is remarkable; see for example Supplementary Material Fig. 3. There is little evidence of debonding between graphite layers; the struts maintain their integrity and do not fragment.

Nominal stress-strain responses of FG and $Al_2O_3$/G foams are compared in Fig. 6a and 6b. Both the FG and $Al_2O_3$/G foams display similar strain hardening behaviors, each exhibiting a plateau in stress between the yield point and densification point $\varepsilon_Y < \varepsilon < \varepsilon_D$ (Fig. 6a). Note that the Al$_2$O$_3$/G foam does not exhibit catastrophic brittle failure, consistent with the fact that ~50 nm thick Al$_2$O$_3$ films are able to sustain small bending radii [54]. Recall from Gibson and Ashby [17] that a brittle foam exhibits a characteristic jagged stress versus strain curve compared to the smooth curves of Fig 6a. For FG foams, the transition from the elastic regime I to the plastic regime II occurs at a yield strain $\varepsilon_Y$ = 0.17–0.40, depending on the magnitude of the relative density $\bar{\rho}$ (in the range 0.002 to 0.005). In contrast, for the Al$_2$O$_3$/G foams, yields occurs at $\varepsilon_Y$ = 0.08–0.21, again depending on the density of the sample measured.

A linear fit to the log-log plots of Figs. 6c and 6d was performed. The slope of the E versus ρ plot has a best fit value of 1.99 (with a 95% confidence interval of 1.66 and 2.32). Similarly, for the $\sigma_Y$ versus ρ plot the best fit slope is 1.32 (with a 95% confidence interval of 1.12 and 1.53). Recall that Gibson and Ashby [17] show that the scaling law reads E $\propto$ $\rho^2$ for cell-wall bending and E $\propto$ ρ for cell-wall stretching. Similarly, the correlation between $\sigma_Y$ and ρ reads $\sigma_Y \propto \rho^{3/2}$ for cell wall bending and $\sigma_Y \propto \rho$ for cell-wall stretching. Taken together, the data of Figs. 6c and 6d support the conclusion that these materials behave as bending-dominated open-cell foams.

It is important to distinguish between the macroscopic density ρ of a foam, when treating it as a homogeneous solid, and the density $\rho_s$ of the cell wall material. For the monolithic free standing graphitic foam, the relative density is $\bar{\rho} = \rho/\rho_s$. Note that $\bar{\rho}$ is identical to the volume fraction of cell wall material in the foam. In contrast, for the composite case of an Al$_2$O$_3$/G foam, it is straightforward to measure ρ, but more involved to determine $\rho_s$ as due account must be made for the proportion of Al$_2$O$_3$ versus graphite. The scaling laws of Gibson-Ashby were established in terms of $\bar{\rho}$ for a monolithic foam and the power-law index is unchanged when strength or modulus is plotted in terms of $\bar{\rho}$ rather than $\rho$.

## 4. Discussion and Modelling

*4.1. Graphitic foam wall thickness and structure*

We adopt the micromechanical Gibson-Ashby approach [17] for bending-dominated open-cell foams in order to interpret the response of the FG and $Al_2O_3$/G foams. The foams in this study are idealized by unit cells with hollow struts of triangular cross-section. The struts have a length *L*, an equivalent side length *d*, and a wall thickness *h*, see Fig. 7a,b. The observed dependence of modulus and yield strength of the FG and $Al_2O_3$/G foam in Fig. 6c,d reveals that $E \propto \bar{\rho}^2$ and $\sigma_{ys} \propto \bar{\rho}^{3/2}$, consistent with strut bending, as anticipated for 3D open-cell foams and lattices of low nodal connectivity [25]. The previous literature on graphene/graphite foams [36,58] has assumed that the pre-factors of the Gibson-Ashby power-law scaling relations [17] are the same as those for metallic and polymeric open-cell foams. However, the deformation mechanisms for the struts of a graphitic foam are much more complex than those of solid struts. The cell walls are hollow, and are made from a layered graphitic structure of low shear modulus and strength.

We utilize a hierarchical micromechanical model spanning three distinct length scales to interpret the mechanical response of the foams in this study. The length scale of the unit cell of the foam is determined by the length of the struts comprising the cell wall, and is termed level I. The cell walls comprise hollow triangular tubes, and the bending curvature of these strut-like tubes involves axial stretching of the tube walls, and this length scale is termed level II. In turn, the tube walls form a wavy stack of graphitic layers, and this waviness induces interlayer shear of the graphitic layers when the tube walls are subjected to axial stretch. The thickness of the tube wall defines the third length scale, termed level III. We emphasize that the hierarchical model of the present study is an idealization to highlight the significance of the microstructure on three length scales.

The properties of the bulk solid, the connectivity and shape of cell edges and faces, and the relative density $\bar{\rho}$ of a cellular solid are the main features that influence cellular properties [18]. Simple scaling laws have been previously derived for idealized cell geometries:

$$\frac{E}{E_s} = \alpha \bar{\rho}^n \qquad (1)$$

$$\frac{\sigma}{\sigma_{ys}} = \beta \bar{\rho}^m \qquad (2)$$

where the relative density $\bar{\rho}$ is the macroscopic apparent density of the foam divided by the density of the constituent solid material. The exponents n and m reflect the deformation mode of the struts within the foam [18,25], and the observed values of n = 2, m = 3/2 are indicative of strut bending behavior. The pre-factors $\alpha$ and $\beta$ depend upon the details of the microstructure [24,59,60]. These scaling laws adequately describe the macroscopic foam behavior for many types of macrocellular foams [61], including ceramics, metals and polymers [62–65]. We shall make use of these power laws in order to interpret the response of the graphitic foams of the present study. We find that the measured values for the pre-factors $\alpha$ and $\beta$ of Eq. 1 and 2 respectively are $7.8 \times 10^{-4}$ and $6.5 \times 10^{-5}$, see Fig. 6c and 6d. These differ greatly from the previously assumed magnitude of pre-factors $\alpha = 1$ and $\beta = 0.3$, as taken from the literature for metallic or polymeric open-cell foams [18,25]. This motivates an investigation into the influence of hollow struts and wavy anisotropic cell walls on the values for the pre-factors ($\alpha, \beta$).

It is recognized that (non-layered) ceramic nano-lattices deform elastically and recover upon unloading [66]. This contrasts with the observed behavior of the multi-layered graphite. We further note that the graphitic foams of the present study deform in a different manner to that of elastic-brittle ceramic foams, see for example Gibson and Ashby [25]. Such foams display a highly jagged stress versus strain response associated with the sequential fracture of individual struts at the loading plateau. No such fragmentation of the struts is observed in the present study. Thus, there is no need to account for fracture energy (such as surface energy) in the hierarchical model.

*4.2. The role of hollow struts*

We first investigate the implications of hollow struts on the stiffness and yield strength, by using the concept of shape factors (see Supplementary Material, Section 3 for details) [67]. The shape factor is a multiplicative scaling factor which expresses the amplification of a mechanical property (such as mechanical modulus), due to a choice of geometry. This factor is normalized by that of a solid circular beam of equal cross-sectional area to that of the geometry under consideration. Shape factors must be taken into consideration to account for this discrepancy between the measured values of the pre-factors $\alpha$ and $\beta$ of the current study and the standard values of $\alpha = 1$ and $\beta = 0.3$, as derived for open-cell foams [25].

Consider the case of an open-cell foam, with cell walls in the form of hollow triangular tubes of wall thickness *h*, strut side length *d* and internal strut side length $d_i$, see Fig. 7b. Assume that the cell walls are made from a solid of Young's modulus $E_s$ and yield strength $\sigma_{ys}$. We further assume the strut to be bending under an applied moment caused by the macroscopic compression of the foam. A reference cell wall of solid circular cross-section of diameter D is used, of cross-sectional area equal to that of the hollow tube, implying $D^2 = 12dh/\pi$. Then, the bending stiffness of the hollow triangular tube equals $\phi_{Be}$ times that of the solid circular bar, such that

$$\phi_{Be} = \frac{2\pi}{27}\frac{d}{h} \quad (3)$$

thereby defining the relevant shape factor for elastic bending of the cell wall struts in the form of hollow tubes.

Next, consider the plastic collapse of a hollow triangular bar and of the solid circular bar of equal cross-sectional area. Upon noting that the plastic collapse moment of the hollow triangular bar $M_{Ph}$ and solid circular bar $M_{Ps}$ are given by $M_{Ph} = \sqrt{3}hd^2\sigma_{ys}$ and $M_{Ps} = D^3\sigma_{ys}/6$ respectively, the relevant shape factor reads

$$\phi_{By} = \frac{\pi\sqrt{\pi}}{4}\left(\frac{d}{h}\right)^{1/2} \quad (4)$$

A direct comparison with Eq. 1 and 2 implies that $\alpha = \phi_{Be}$ and $\beta = 0.3\phi_{By}$. Using values of wall thickness h and strut width d, as measured by cross-sectional SEM, we determine the value of the shape factor for elastic bending to be $\phi_{Be} \approx 80$, and the shape factor for failure in bending $\phi_{By} \approx 30$. Shape factors exceeding unity are typical of hollow sections [68]. However, these values are several orders of magnitude too large when compared to the experimentally determined values for the constants of proportionality, indicating that the high-aspect ratio cross-sectional shape alone cannot account for the significant reduction in stiffness and strength. We seek an explanation at a lower length scale, that of the walls of the hollow triangular struts.

*4.3. Effect of wall waviness*

We emphasize that the above calculation of shape factors for a hollow triangular beam is based on the assumption that the walls of the hollow cross-section are perfectly straight. In reality the walls are wavy, as demonstrated by the high-resolution SEM images in Fig. 3 and Supplementary Material Fig. 6. The walls of the hollow tubes are subjected to a gradient of axial stress from tension in the top fiber to compression in the bottom fiber when the tube is subjected to a bending moment M. Recall that these walls comprise a multi-layered stack of graphitic sheets, see Figs. 1, 2. When this wavy stack of sheets is subjected to an axial tension or compression, this misalignment induces bending loads and transverse shear forces on the cross-section of the cell wall. The wavy sheet responds by bending and by shear deflections, which lead to a change in the axial length of the wavy stack of sheets.

*4.3.1. Wall bending*

We idealize the waviness by a sine wave of amplitude $w_0$ and wavelength of $\lambda$, such that the transverse deflection in the initial, unloaded state is

$$w(x) = w_0 sin\left(\frac{2\pi x}{\lambda}\right) \qquad (5)$$

The axial compliance of each face of the triangular strut is increased due to this waviness. Consequently, bending due to this waviness will introduce a knock-down factor $k_{Be}$ in the effective modulus of the cell walls and also in the macroscopic modulus of the foam, as well as a reduction in the axial strength by a knock-down factor $k_{By}$. The magnitude of these knock-down factors are obtained by treating the cell wall as a beam of height h and assuming that the axial straining of a wavy beam is driven by an end tension T, as depicted in Fig. 7c. Under an end tension T, the beam bends locally due to a bending moment $M = Tw$, and consequently the beam straightens and lengthens. Elementary beam bending theory (see Supplementary Material, Section 4.1 for details) suggests that the initial waviness reduces the axial stiffness of the wavy beam compared to that of a straight beam by a scale factor, and gives us:

$$k_{Be} = \frac{1}{6}\left(\frac{h}{w_0}\right)^2 \tag{6}$$

Similarly, the axial strength of a wavy beam is less than that of the equivalent straight beam due to waviness inducing local bending within the beam, so that it undergoes plastic collapse by hinge formation at the location of maximum waviness. The knock down factor in yield strength due to the waviness is given by,

$$k_{By} = \frac{h}{4w_0} \tag{7}$$

The macroscopic modulus of an elastic foam, upon neglecting correction factors, is given by

$$E = \bar{\rho}^2 E_s \tag{8}$$

when cell wall bending dominates the response, that is $\alpha = 1$ and $n = 2$, as discussed by Gibson & Ashby [25]. Now modify Eq. 8 by the presence of the shape factor $\phi_{Be}$ and the knockdown factor $k_{Be}$ at two structural hierarchies, such that

$$E = \phi_{Be} k_{Be} \bar{\rho}^2 E_s \tag{9}$$

This is of the form of Eq. 1 but with a correction pre-factor $\alpha$ now given by

$$\alpha = \phi_{Be} k_{Be} \qquad (10)$$

Likewise (see Supplementary Material, Section 4.2 for details), the macroscopic yield strength for bending-dominated open-cell foams, absent any correction factor, is

$$\sigma_y = 0.3 \bar{\rho}^{3/2} \sigma_{ys} \qquad (11)$$

implying $\beta = 0.3$ and $m = 3/2$. Now modify Eq. 11 by the presence of the shape factor $\phi_{By}$ and the knockdown factor $k_{By}$ at two structural hierarchies, such that

$$\sigma_y = 0.3 \phi_{By} k_{By} \bar{\rho}^{3/2} \sigma_{ys} \qquad (12)$$

This is of the form of Eq. 2 but with a correction pre-factor $\beta$ given by

$$\beta = 0.3 \phi_{By} k_{By} \qquad (13)$$

*4.3.2. Wall shear*

Cell wall waviness can induce an alternative deformation mechanism, that of cell-wall shear. The wavy multilayer walls of the hollow triangular struts undergo shear loading when the faces of the struts are loaded by axial stress. Recall that these axial stresses arise from bending of the cell walls of the open-cell foam. Consequently, the waviness gives a knock-down factor $k_{Se}$ in the macroscopic modulus and a knockdown factor $k_{Sy}$ in the macroscopic yield strength of the foam.

Consider a wavy face of the triangular tube with a shear modulus $G_s$ and shear strength $\tau_{ys}$. Then, the axial stiffness of the wavy beam of thickness h is knocked-down from that of the equivalent straight beam by a factor $k_{Se}$, and likewise the axial strength is knocked down by a factor $k_{Sy}$, where elementary beam theory (see Supplementary Material, Section 4.3 and 4.4 for details) gives

$$k_{Se} = \frac{1}{2}\left(\frac{\lambda}{\pi w_0}\right)^2 \frac{G_s}{E_s} \tag{14}$$

$$k_{Sy} = \frac{1}{2\pi} \frac{\lambda}{w_0} \frac{\tau_{ys}}{\sigma_{ys}} \tag{15}$$

The relation between macroscopic foam modulus $E$ and yield stress $\sigma_y$ due to cell wall shear follows from insertion of Eq. 14 into 8 to give

$$E = \phi_{Be} k_{Se} \bar{\rho}^2 E_s \tag{16}$$

implying that

$$\alpha = \phi_{Be} k_{Se} \tag{17}$$

Likewise, the yield strength of the foam now reads

$$\sigma_y = 0.3 \phi_{By} k_{Sy} \bar{\rho}^{3/2} \sigma_{ys} \tag{18}$$

implying

$$\beta = 0.3 \phi_{By} k_{Sy} \tag{19}$$

*4.4. Failure modes*

We emphasize that the multi-scale model assumes that the knock-down factors at each length scale act independently of each other. This is reasonable when there is a wide separation of length scales, as in the present study. Accordingly, the overall knock-down factor is determined by the product of knock-down factors at each length scale. It is clear from Eq. 6 to 18 that there exists a strong dependence of macroscopic modulus and strength on the amplitude of the wall waviness $w_0$.

In order to assess which failure mode is active, waviness amplitude values were determined that are in agreement with measured values of macroscopic modulus and strength, assuming that the hollow cell walls of the foam undergo either bending or shear. Predictions of the amplitude of waviness $w_0$ are obtained from equations (10), (13), (17) and (19) (see Supplementary Material Section 5). We find that wall shear implies waviness amplitudes in the range of 0.45 μm to 22 μm, whereas hollow wall bending calls for waviness amplitudes of 11 μm to 5800 μm. (Table 2). SEM images of edge profiles of the graphitic struts reveal waviness amplitudes on the order of 2.8 μm, indicating that the deformation of multi-layer graphitic foams is dominated by interlayer shear rather than intralayer bending. This is consistent with the large contrast between the high in-plane Young's modulus and low out-of-plane shear modulus of multi-layer graphene [69].

The above hierarchical model uses the language of plasticity theory, with the notion that bending of the struts is by plastic slip between planes of the graphitic walls. This is consistent with observations on the nanoscale of the deformation of CVD-grown graphitic layers in cantilever beams [70]. Carbon nanotubes also display similar behavior with longitudinal plastic shear between the layers of a nanotube [71,72]. Compared to other 3D graphene-based assemblies [1,2,9,73–75], uniaxial compression studies on graphene-based aerogels have observed yield strength scaling as $\bar{\rho}^{2.4}$ (see Supplementary Material Fig. 7). Recall that $\sigma_y \propto \bar{\rho}^{3/2}$ in the present study. The discrepancy between values for the exponent can be traced to the fact that aerogel foams comprise a percolating network of stacked graphitic platelets, rather than the continuously grown sheets that form the foams in our study that afford a more electrically conductive networked structure (see Supplementary Material Fig. 8).

## 5. Concluding Remarks

The compressive response of freestanding CVD graphitic foams has been measured for a range of relative densities, and a three level hierarchical model has been developed to explain the dependence of modulus and strength upon relative density and microstructure. As the basis for a reproducible model system we used commercial Ni templates and a graphitic wall thickness larger than 80 nm in combination with process and handling improvements such as $H_2$ annealing and laser sectioning.

The power law dependence of compressive modulus and yield strength of the open-cell foam suggests that the cell walls undergo beam bending (level I). However, the measured pre-factors in the power laws are several orders of magnitude lower than those observed for conventional polymeric and metallic open-cell foams. This knock-down is traced to the following microstructural features. The cell struts are hollow tubes (level II), with wavy walls, and consequently the axial stiffness and strength of the faces of the tube are degraded by the waviness (level III). By comparing predicted levels of waviness with measured values, we have demonstrated that the dominant failure mechanism is inter-layer shear rather than in-plane bending of the wavy walls. These factors lead to a multiplicative knock-down in macroscopic properties.

We have also explored the addition of a thin, flexible ceramic ALD $Al_2O_3$ scaffold to the freestanding graphitic foams. There is an increasing body of literature to suggest that ultrathin ceramic metamaterials exhibit ductile behavior when wall thicknesses fall below 100 nm [54,66,76,77]. The results of the present study are consistent with these findings; the graphitic foams tested herein possess a cell wall thickness on the order of 80-150 nm, with a 50 nm thick alumina scaffold. We found this thin ceramic scaffold increases the strength and stiffness of the foams while still conforming to the same scaling laws as those exhibited by the freestanding graphitic foams. The micromechanical, hierarchical model presented here represents a first step towards an understanding of graphitic foams across multiple length scales. Additionally, our findings suggest future research directions for the design of 2D material-based cellular materials and their emergent applications.

**Acknowledgements**

We acknowledge funding from EPSRC (Grant No. EP/K016636/1, GRAPHTED) and the ERC (Grant No. 279342, InsituNANO; Grant No. 669764, MULTILAT). A.I.A. acknowledges the Green Talents Research Stay program from The German Federal Ministry of Education and Research (BMBF). K.N. acknowledges funding from the EPSRC Cambridge NanoDTC (Grant No. EP/G037221/1).

**Table 1.** Summary of key measured length-scales as measured by cross-sectional SEM (Fig 3 and Supplementary Material Fig. 6).

| Length-scale | h (μm) | d (μm) | $w_0$ (μm) | λ (μm) |
|---|---|---|---|---|
| **Minimum** | 0.08 | 35 | 0.76 | 3.7 |
| **Maximum** | 0.20 | 65 | 2.8 | 18 |

**Table 2.** Summary of predicted waviness amplitudes for wall bending vs. wall shearing elastic and yield behavior.

| Scenario | Predicted $w_0$ (μm) |
|---|---|
| **Elastic Wall Bending (eq. 9)** | $11 - 26$ |
| **Plastic Wall Bending (eq. 12)** | $2700 - 5800$ |
| **Elastic Wall Shear (eq. 16)** | $2.1 - 22$ |
| **Plastic Wall Shear (eq. 18)** | $0.45 - 4.8$ |

**Figures**

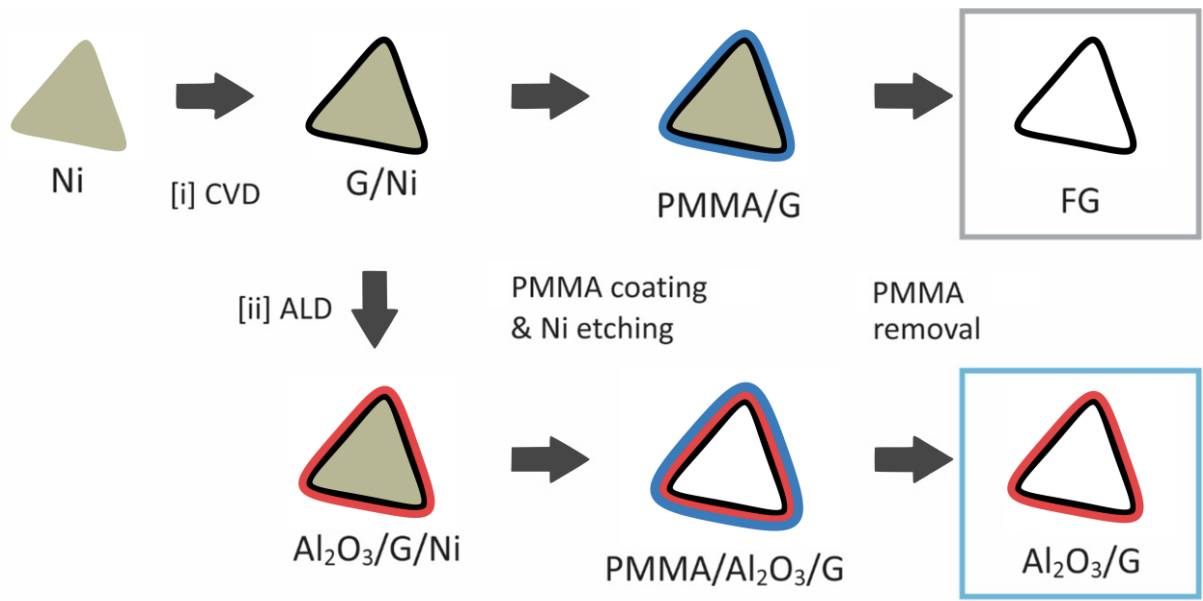

Figure 1. Schematic of the synthesis of freestanding graphene (FG) and ceramic composite graphitic ($Al_2O_3$/G) foams. [i] Nickel foams (Ni) are used as the templates on which graphene layers are grown (G/Ni) by CVD. [ii] For the fabrication of $Al_2O_3$/G, the as-grown G/Ni are coated with $Al_2O_3$ ($Al_2O_3$/G/Ni) film by means of ALD. A PMMA layer is then coated onto these foams to provide structural support during the removal of Ni templates by means of wet chemical etching. FG and $Al_2O_3$/G foams are then obtained once the PMMA coating has been removed by $H_2$ annealing.

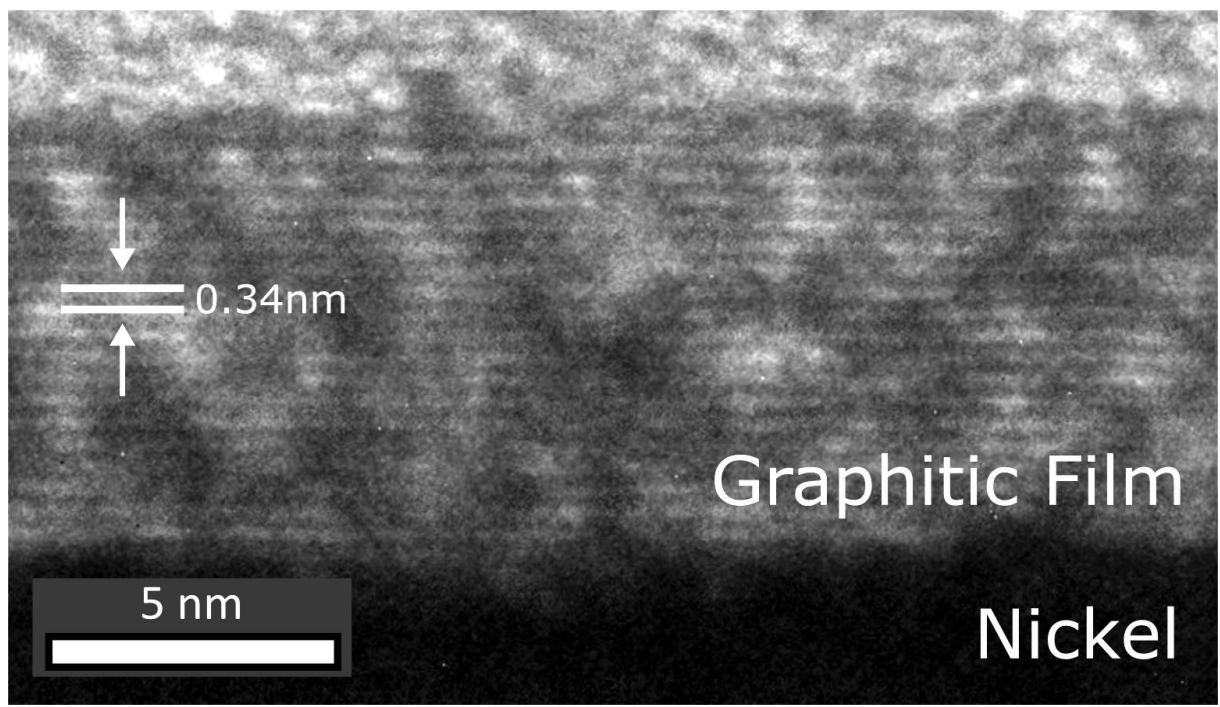

Figure 2. HR-TEM of a CVD-deposited few-layer graphene on a Ni template, showing conformal growth of stacked layers. The lattice image taken from a surface cross-section of a CVD-deposited graphitic film on a catalytic film shows graphitic layers running parallel to the metal surface with characteristic (002) graphite spacing. The sample was grown [∼1000 °C, $CH_4$(10 sccm)/$H_2$(600 sccm), 3 min, cooled at ∼25 °C/min] on a Ni-Au film (550 nm thick, 1.2% Au alloy), with the Au admixture giving improved nucleation control.

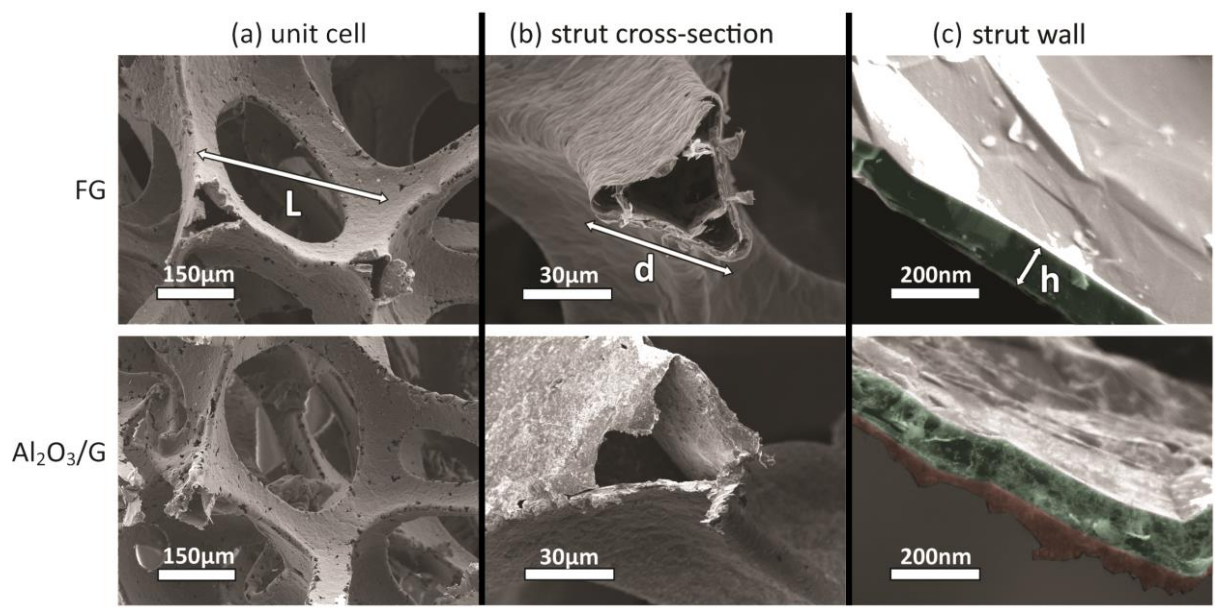

Figure 3. SEM images of FG and $Al_2O_3$/G at different magnifications showing their typical (a) unit cell, (b) strut cross-section, and (c) strut wall. (a) The cellular geometries of FG and $Al_2O_3$/G closely resemble those of Ni foam templates with approximate unit cell length (L) of 300(±100) μm. (b) A cross-sectional cut shows that the struts of both FG and $Al_2O_3$/G are hollow with triangular cross-section and equivalent side length (d) of 50(±15) μm. (c) The strut wall of FG consists of hundreds of graphene layers, as highlighted in green, with thickness h that varies between 80 nm and 150 nm. The strut wall of $Al_2O_3$/G consists of graphene layers and $Al_2O_3$ film, as highlighted in green and red, respectively, with an overall thickness (h) that varies between 130 and 200 nm

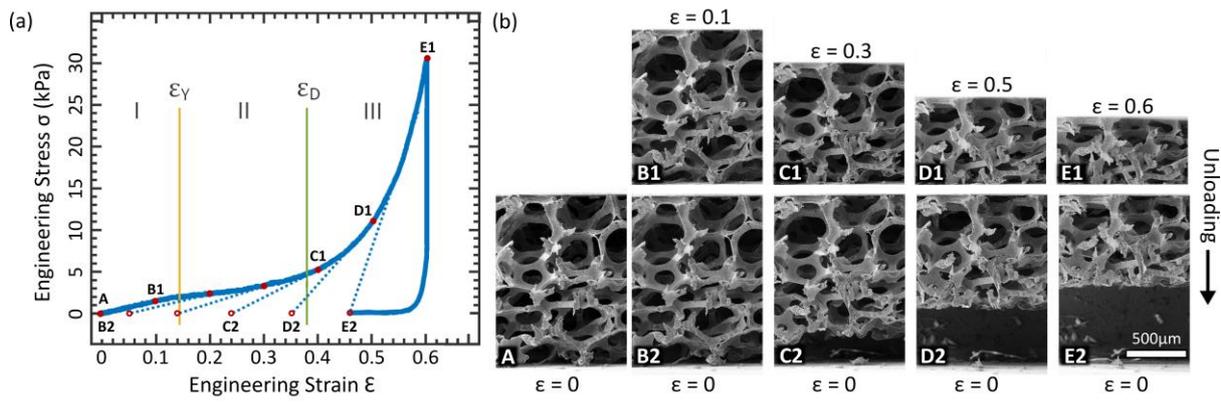

Figure 4. (a) Typical nominal stress-strain curve response of FG under compressive load. Three distinct regimes exist: (I) linear elastic $\varepsilon < \varepsilon_Y$, (II) plateau $\varepsilon_Y < \varepsilon < \varepsilon_D$, and (III) densification $\varepsilon > \varepsilon_D$. (b) SEM images of a sample subjected to successively larger levels of macroscopic strain $\varepsilon$, followed by unloading to zero load. The remnant strain $\varepsilon_r$ was measured from the associated SEM images. A dotted line is drawn to illustrate this elastic unloading.

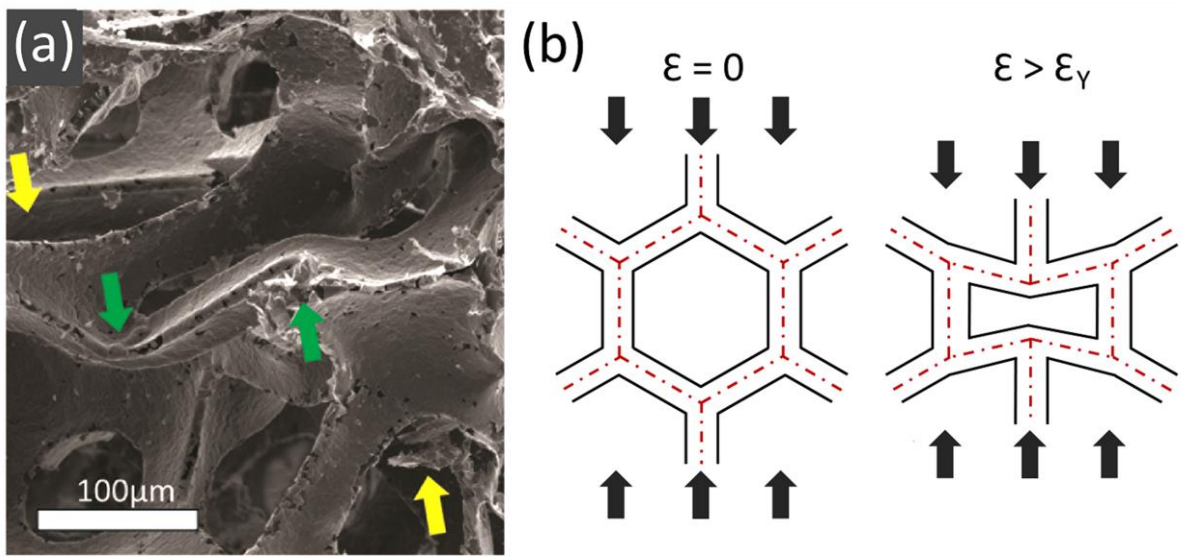

Figure 5. (a) Low magnification SEM image of FG foam taken post compression (peak $\varepsilon$ = 0.6). Green and red arrows indicate the plastically deformed and fractured struts, respectively. The green arrows indicate a pair of plastic hinges formed on the deformed strut. The yellow arrows indicate the direction of the compressive force. (b) Schematic of the plastically deformed strut at $\varepsilon > \varepsilon_Y$, with the cell structure skeleton outlined in red. The struts undergo bending when the foam is subjected to compressive force. Since the strut ends are rigid and act as rotation-fixed but translation-free constraints, a pair of plastic hinges is formed on the strut.

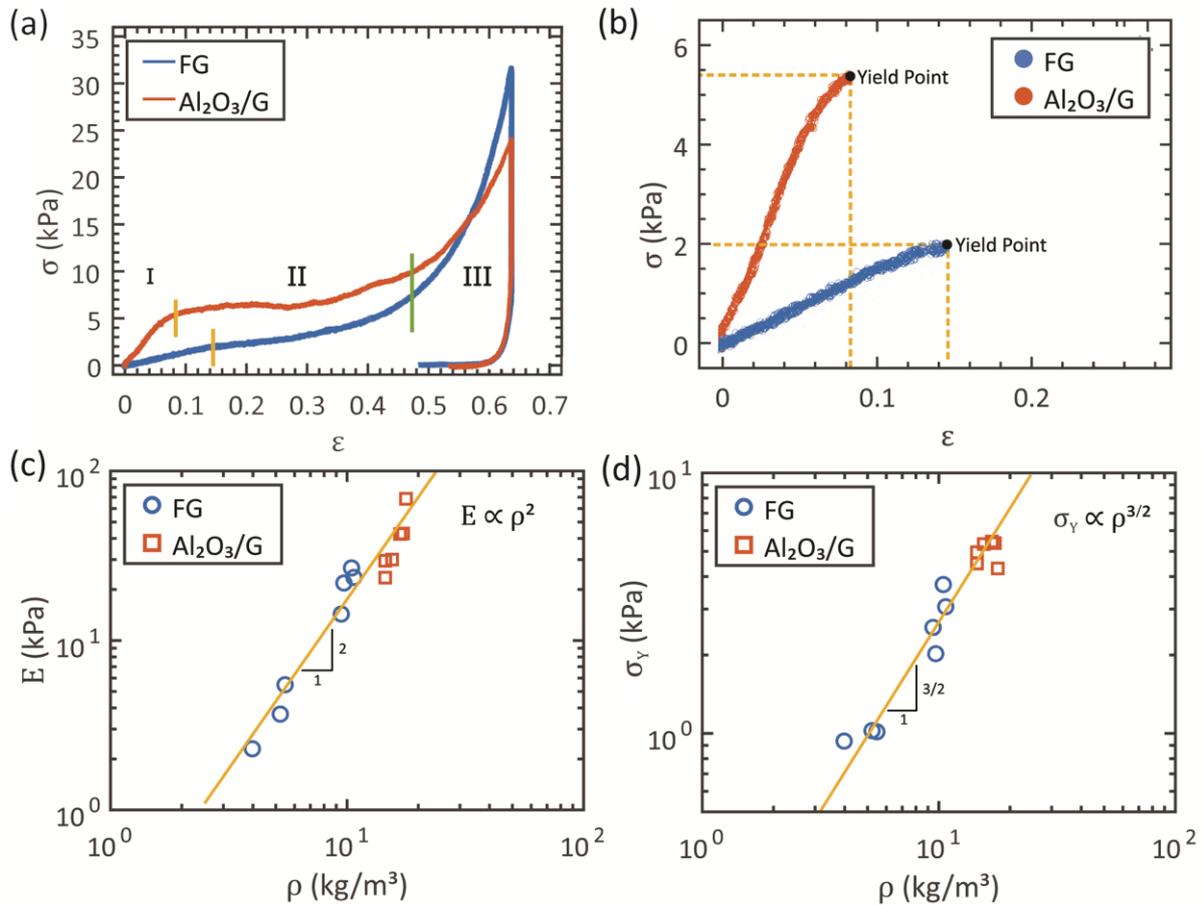

Figure 6. (a) Typical nominal stress-strain (σ-ε) response of FG and $Al_2O_3$/G under compressive load with a displacement rate of ~10μm/s. The onset of plasticity ε = $ε_Y$, is indicated by the yellow line, while the onset of densification ε = $ε_D$, is indicated by the green line. (b) Sample σ-ε curves of FG and $Al_2O_3$/G in the linear elastic regime. The yield stress ($σ_Y$) is given by the stress at $ε_Y$, while the compressive modulus (E) is obtained by a linear fit. Plot of E (c) and $σ_Y$ (d) of FG and $Al_2O_3$/G as a function of their apparent density (ρ). The yellow fit in (c) and (d) indicate scaling of $E \propto \rho^2$ and $σ_Y \propto \rho^{3/2}$, respectively.

## (a) Cell Structure

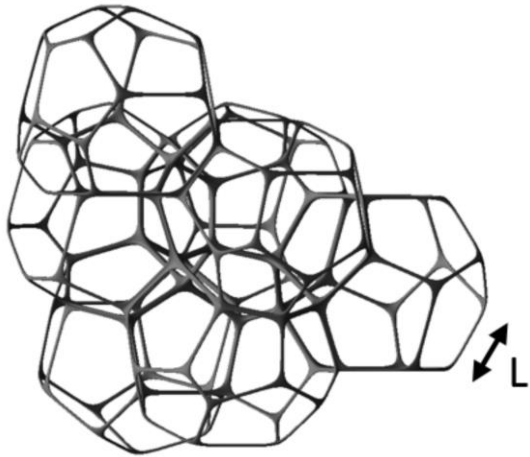

## (b) Strut Cross-section

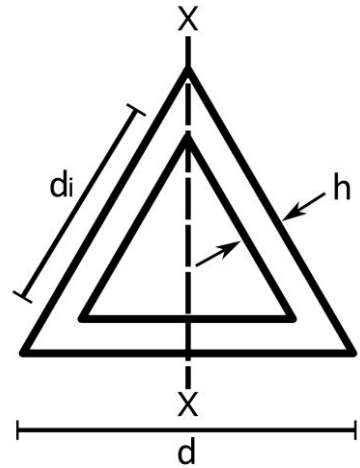

## (c) Wavy Wall

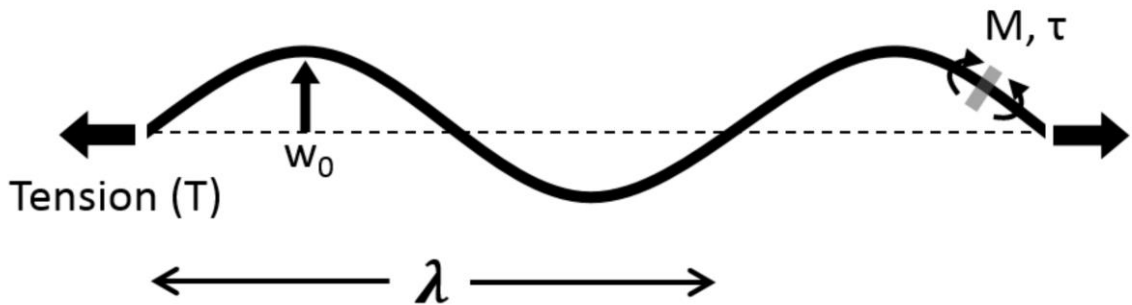

Figure 7. FG and $Al_2O_3$/G structure idealization. (a) The Weaire-Phelan open-cell foam structure, a bending-dominated, idealized foam of cells with equal volume. For the foams studied herein, the struts are hollow with an approximate length of L. (b) Strut schematic illustrating the hollow triangular cross-section with a side length of d, internal side length $d_i$, bending axis X--X and a wall thickness of h. (c) Wall level schematic of a cell wall loading in the micromechanical models used herein. The wall waviness is represented as a sine wave of amplitude $w_0$ and wavelength λ. In a wavy wall subjected to an axial tension or compression, misalignment induces bending loads and transverse shear forces on the cross-section of the cell wall, leading to the suggested wall bending or wall shear deformation modes.